\documentclass[12pt]{article}
\usepackage{amssymb,amsmath,color,natbib,graphicx,amsthm,
  setspace,sectsty,anysize,times,dsfont}

\usepackage{lscape,arydshln,relsize,rotating}
\usepackage[small]{caption}

\marginsize{1.1in}{.9in}{.3in}{1.4in}

\newcommand{\dbl}{\setstretch{1.5}}
\newcommand{\sgl}{\setstretch{1.1}}

\newcommand{\bs}[1]{\boldsymbol{#1}}
\newcommand{\mc}[1]{\mathcal{#1}}
\newcommand{\mr}[1]{\mathrm{#1}}
\newcommand{\bm}[1]{\mathbf{#1}}
\newcommand{\ds}[1]{\mathds{#1}}
\newcommand{\indep}{\perp\!\!\!\perp}

\newcommand{\cd}{\smaller\sf}
\newcommand{\e}[1]{{\footnotesize$\times10$}{$^{#1}$}}

\sectionfont{\noindent\normalfont\large\bf}
\subsectionfont{\noindent\normalfont\normalsize\bf}
\subsubsectionfont{\noindent\normalfont\it}

\begin{document}

\sgl

\pagestyle{empty}

~
\vskip 2cm

\noindent {\Large \bf{Measuring political sentiment on Twitter:  

    \vskip .2cm \hskip .25cm factor-optimal design for multinomial
    inverse regression}}
\vskip 1cm

\noindent{\large Matt Taddy}\\
{\tt taddy@chicagobooth.edu}

\vskip .2cm
\noindent{The  University of Chicago Booth School of Business}\\
{5807 South Woodlawn, Chicago IL 60637}

\vskip 3cm
\noindent {\small {\bf Keywords} 

\noindent   text analysis, sentiment mining,
  inverse regression, multinomial logistic regression,\\ topic models, optimal
  design, active learning, variable interaction}

\vskip .5cm

{\small
  \noindent {\bf Abstract} 

\noindent  This article presents a short case study
  in text analysis: the scoring of Twitter posts for positive,
  negative, or neutral sentiment directed towards particular US
  politicians.  The study requires selection of a sub-sample of
  representative posts for sentiment scoring, a common and costly
  aspect of sentiment mining. As a general contribution, our
  application is preceded by a proposed algorithm for maximizing
  sampling efficiency.  In particular, we outline and illustrate
  greedy selection of documents to build designs that are D-optimal in
  a topic-factor decomposition of the original text.  The strategy is
  applied to our motivating dataset of political posts, and we outline
  a new technique for predicting both generic and subject-specific
  document sentiment through use of variable interactions in
  multinomial inverse regression.  Results are presented for analysis
  of 2.1 million Twitter posts collected around February 2012.  Computer codes
  and data are provided as supplementary material online. }

\newpage
\dbl

\pagestyle{plain}
\vskip 1cm
\section{Introduction}
\label{intro}

This article outlines a simple approach to a general problem in text
analysis, the selection of documents for costly annotation. We then
show how inverse regression can be applied with variable interactions
to obtain both generic and subject-specific predictions of document
sentiment, our annotation of interest.  
We are motivated by the problem of design and analysis of
a particular text mining experiment: the scoring of Twitter posts (`tweets') for
positive, negative, or neutral sentiment directed towards particular
US politicians.  The contribution is structured first with a proposal
for optimal design of text data experiments, followed by application
of this technique in our political tweet case study and analysis of
the resulting data through inverse regression.

Text data are viewed throughout simply as counts, for each document,
of phrase occurrences.  These phrases can be words (e.g., {\it tax})
or word combinations (e.g. {\it pay tax} or {\it too much tax}).
Although there are many different ways to process raw text into these
{\it tokens}, perhaps using sophisticated syntactic or semantic rules,
we do not consider the issue in detail and assume tokenization as
given; our case study text processing follows a few simple rules
described below.  Document $i$ is represented as $\bm{x}_i =
[x_{i1},\ldots,x_{ip}]'$, a sparse vector of counts for each of $p$
tokens in the vocabulary, and a document-term count matrix is written
$\bm{X} = [\bm{x}_1 \cdots \bm{x}_n]'$, where $n$ is the number of
documents in a given corpus. These counts, and the associated
frequencies $\bm{f}_i = \bm{x}_i/m_i$ where $m_i = \sum_{j=1}^p
x_{ij}$, are then the basic data units for statistical text analysis.
Hence, text data can be characterized simply as exchangeable counts in
a very large number of categories, leading to the common assumption of
a multinomial distribution for each $\bm{x}_i$.

We are concerned with predicting the {\it sentiment} $\bm{y} =
[y_1,\ldots,y_n]'$ associated with documents in a corpus.  In our main
application, this is positive, neutral, or negative  sentiment
directed toward a given politician, as measured through a reader
survey.  More generally, sentiment can be replaced by any annotation
that is correlated with  document text.  Text-sentiment
prediction is thus just a very high-dimensional regression problem,
where the covariates have the special property that they can be
represented as draws from a multinomial distribution.

Any regression model needs to be accompanied with data for training.
In the context of sentiment prediction, this implies documents scored
for sentiment.  One can look to various sources of `automatic'
scoring, and these are useful to obtain the massive amounts of data
necessary to train high-dimensional text models. Section \ref{data}
describes our use of emoticons for this purpose.  However, such
automatic scores are often only a rough substitute for the true
sentiment of interest.  In our case, generic happy/sad sentiment is
not the same as sentiment directed towards a particular politician.
It is then necessary to have a subset of the documents annotated with
precise scores, and since this scoring will cost money we need to
choose a subset of documents whose content is most useful for
predicting sentiment from text.  This is an application for {\it pool
  based active learning}: there is a finite set of examples for which
predictions are to be obtained, and one seeks to choose an optimal
representative subset.

There are thus two main elements to our study: design -- choosing the
sub-sample of tweets to be sent for scoring -- and analysis -- using
sentiment-scored tweets to fit a model for predicting Twitter sentiment
towards specific politicians.  This article is about both components.
As a design problem, text mining presents a difficult situation where
raw space filling is impractical -- the dimension of $\bm{x}$ is so
large that every document is very far apart -- and we argue in Section
\ref{design} that it is unwise to base design choices on the poor
estimates of predictive uncertainty provided by text regression.  Our
solution is to use a space-filling design, but in an estimated lower
dimensional multinomial-factor space rather than in the original
$\bm{x}$-sample. Section \ref{design}.1 describes a standard class of
{\it topic models} that can be used to obtain low-dimensional factor
representations for large document collections.  The resulting
unsupervised algorithm (i.e., sampling proceeds without regard to
sentiment) can be combined with any sentiment prediction model.  We
use the multinomial inverse regression of \cite{Tadd2012a}, with the
addition of politician-specific interaction terms, as described in
Section \ref{mnir}.

\subsection{Data application: political sentiment on Twitter}
\label{data}

\begin{figure}[t]
\includegraphics[width=6.3in]{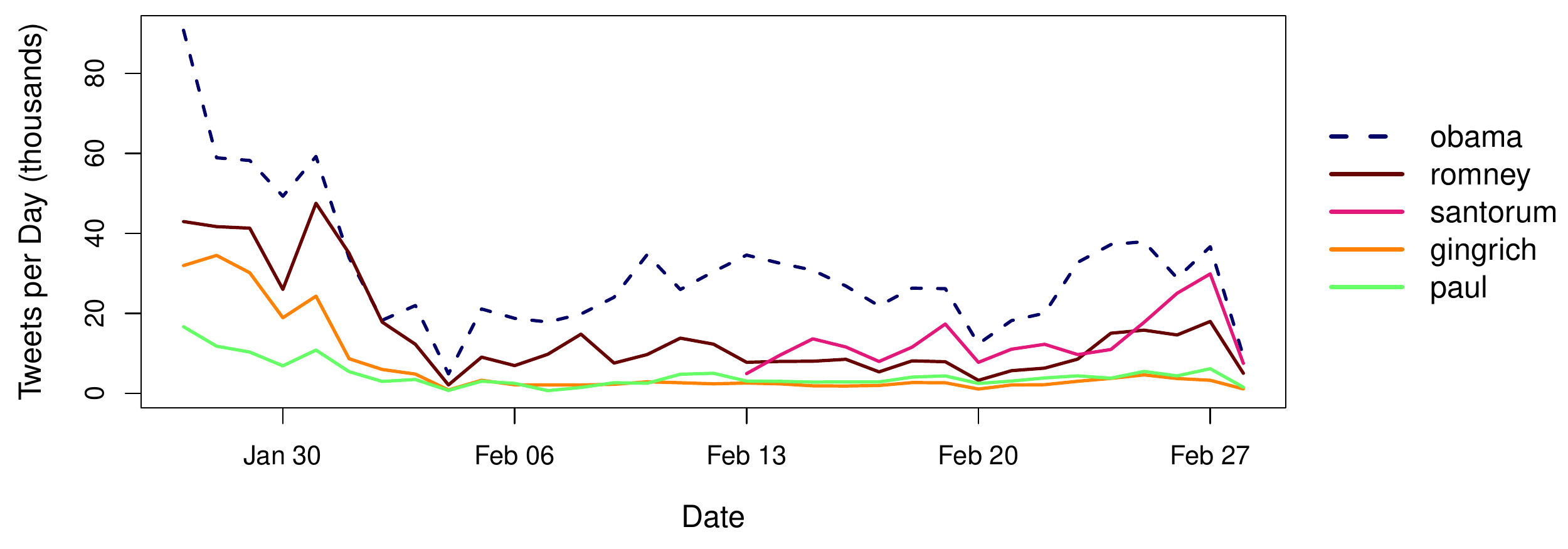}
\caption{\label{volume} Tweet sample volume for political
  candidates. All are taken from the stream of public Twitter posts
  from Jan 27 through the end of February, except for Santorum who was
  only tracked after Feb 13. }
\end{figure}

The motivating case study for this article is an analysis of sentiment
in tweets about US politicians on Twitter, the social blog,
from January 27 to February 28, 2012, a period that included the
Florida (1/31), Nevada (2/4), Colorado, Missouri, and Minnesota (2/7),
Maine (2/11), and Michigan and Arizona (2/28) presidential primary
elections.  Twitter provides streaming access to a large subset of
public (as set by the user) tweets containing terms in a short list of
case insensitive filters.  We were interested in conversation on the
leading candidates in the Republican presidential primary, as well as
that concerning current president Barack Obama; our list of filter
terms was {\cd obama}, {\cd romney}, {\cd gingrich}, {\cd ron paul},
and, from February 13 onward, {\cd santorum}.  Note that Romney,
Gingrich, and Paul were the only front-runners at the beginning of our
study, but Santorum gained rapidly in the polls following his surprise
victories in three state votes on February 7: the Minnesota and
Colorado caucuses and the Missouri Primary.  Daily data collection is
shown by politician-subject in Figure \ref{volume}; total counts are
10.2\e{5} for Obama, 5\e{5} for Romney, 2.2\e{5} for Gingrich,
2.1\e{5} for Santorum, and 1.5\e{5} for Paul, for a full sample of
about 2.1 million tweets.

In processing the raw text, we remove a limited set of stop words
(terms that occur at a constant rate regardless of subject, such as
{\it and} or {\it the}) and punctuation before converting to lowercase
and stripping suffixes from roots according to the Porter stemmer
\citep{Port1980}.  The results are then tokenized into single terms
based upon separating white-space, and we discard any tokens that
occur in $<$ 200 tweets and are not in the list of tokens common in
our generic emoticon-sentiment tweets, described in the next
paragraph.  This leads to 5532 unique tokens for Obama, 5352 for
Romney, 5143 for Gingrich, 5131 for Santorum, and 5071 for Paul.

The primary analysis goal is to classify tweets by
sentiment: positive, negative, or neutral.  We have two data sources
available: twitter data that is scored for generic
sentiment, and the ability to survey readers about sentiment
in tweets directed at specific politicians.  In the first case, 1.6 million
tweets were obtained, from the website {\cd
  http://twittersentiment.appspot.com}, that have been automatically
identified as positive or negative by the presence of an emoticon
(symbols included by the author -- e.g., a happy face indicates a
positive tweet and a sad face a negative tweet).  Tokenization for
these tweets followed the same rules as for the political Twitter
sample above, and we discard tokens that occur in less than 0.01\%
of tweets.  This leads to a vocabulary of 5412 `emoticon' tokens; due
to considerable overlap, the combined vocabulary across all tweets
(political and emoticon) is only 5690 tokens.

As our second data source, we use the Amazon Mechanical Turk ({\cd
  https://www.mturk.com/}) platform for scoring tweet sentiment.
Tweets are shown to anonymous workers for categorization as
representing either positive (e.g., `support, excitement, respect, or
optimism') or negative (e.g., `anger, distrust, disapproval, or
ridicule') feelings or news towards a given politician, or as neutral
if the text is `irrelevant, or not even slightly positive or
negative'.  Each tweet is seen by two independent workers, and it is
only considered scored if the two agree on categorization.  In
addition, workers were pre-screened as `masters' by Amazon and we
monitored submissions for quality control, blocking poor workers.
Given the 2-3 cents per-tweet paid to individual workers, as well as
the overhead charged by Amazon, our worker agreement rates of around
80\% imply an average cost near \$0.075 per sentiment scored tweet.

\section{Sentiment prediction via multinomial inverse regression}
\label{mnir}

Sentiment prediction in this article follows the multinomial inverse
regression (MNIR) framework described in \citet{Tadd2012a}.  
Section \ref{mnir}.1 summarizes that approach, while Section 2.2 
discusses an adaptation specific to the main application of this paper. Inverse
regression as a general strategy looks to estimate the {\it inverse
   distribution} for covariates given response, and to use
this as a tool in building a {\it forward model} for $y_i$ given
$\bm{x}_i$.  The specific idea of MNIR is to estimate a simple model
for how the multinomial distribution on text counts changes with
sentiment, and to derive from this model low dimensional text
projections that can be used for predicting sentiment.

\subsection{Single-factor MNIR}

As a simple case, suppose that $y_i$ for document $i$ is a discrete
ordered sentiment variable with support $\mc{Y}$ -- say $y_i \in
\{-1,0,1\}$ as in our motivating application.  Only a very complicated
model will be able to capture the generative process for an
individual's text, $\bm{x}_i |y_i$, which involves both
heterogeneity between individuals and correlation across dimensions of
$\bm{x}_i$.  Thus estimating a model for $\bm{x}_i |y_i$ can be far
harder than predicting $y_i$ from $\bm{x}_i$, and inverse regression
does not seem a clever place to be starting analysis.  However, we can
instead concentrate on the {\it population average} effect of
sentiment on text by modeling the conditional distribution for
collapsed token counts $\bm{x}_{y} = \sum_{i:y_i=y} \bm{x}_i$.
A basic MNIR model is then
\begin{equation} \label{basic-mnir} \bm{x}_{y} \sim \mr{MN}(\bm{q}_{y},
  m_{y})~~\text{with}~~ q_{yj} = \frac{\exp[\alpha_j +
    y\varphi_j]}{\sum_{l=1}^p \exp[\alpha_l + y\varphi_l
    ]},~~\text{for}~~j=1,\ldots,p,~~y \in \mc{Y}
\end{equation}
where each $\mr{MN}$ is a $p$-dimensional multinomial distribution
with size $m_{y} = \sum_{i:y_i=y} m_i$ and probabilities
$\bm{q}_{ y} = [q_{y1},\ldots,q_{yp}]'$ that are a linear function of
$y$ through a logistic link.  Although independence assumptions
implied by (\ref{basic-mnir}) are surely incorrect, within-individual
correlation in $\bm{x}_i$ is quickly overwhelmed in aggregation and
the multinomial becomes decent model for $\bm{x}_{y}$.  (One
could also argue against an equidistant three point scale for $y$;
however such a scale is useful to simplify inverse regression
and we assume that misspecification here can be accommodated in
forward regression).

Given sentiment $y$ and counts $\bm{x}$ drawn from the multinomial
distribution $\mr{MN}(\bm{q}_{y}, m)$ in (\ref{basic-mnir}), the
projection $\bs{\varphi}'\bm{x}$ is {\it sufficient for sentiment} in
the sense that $y \indep \bm{x} \mid \bs{\varphi}'\bm{x}, m$.  A
simple way to demonstrate this is through application of Bayes rule
(after assigning prior probabilities for each element of $\mc{Y}$).
Then given $\bm{x}_i$ counts for an {\it individual} document,
$\bs{\varphi}'\bm{x}_i$ seems potentially useful as a low-dimensional
index for predicting $y_i$.  More specifically, we normalize by
document length in defining the {\it sufficient reduction} (SR) score
\begin{equation}\label{basic-sr}
z_i = \bs{\varphi}'\bm{f}_i = \bs{\varphi}'\bm{x}_i/m_i.
\end{equation}

Now, since (\ref{basic-mnir}) is a model for collapsed text counts
rather than for $\bm{x}_i$ given $y_i$, the SR score in
(\ref{basic-sr}) is {\it not} theoretically sufficient for that
document's sentiment.  \cite{Tadd2012a} describes specific random
effects models for the information loss in regressing $y_i$ onto $z_i$
instead of $\bm{x}_i$, and under certain models the individual
document regression coefficients approach $\bs{\varphi}$.  However, in
general this population average projection is {\it misspecified} as an
individual document projection.  Hence, instead of applying Bayes rule
to invert (\ref{basic-mnir}) for sentiment prediction, $z_i$ is
treated as an observable in a second-stage regression for $y_i$ given
$z_i$.  Throughout this article, where $y$ is always an ordered
discrete sentiment variable, this {\it forward regression} applies
logistic proportional odds models of the form $\mr{p}(y_i < c) =
\left(1 + \exp[ -(\gamma_c + \beta z_i)]\right)^{-1}$.

\subsection{MNIR with politician-interaction}

In the political twitter application, our approach needs to be adapted
to allow different text-sentiment regression models for each
politician, and also to accommodate positive and negative emoticon tweets, 
which are sampled from all public tweets rather than always being associated 
with a politician.  This is achieved naturally within the MNIR framework by
introducing interaction terms in the inverse regression.

The data are now written with text in the i$^{th}$ tweet for
politician $s$ as $\bm{x}_{si}$, containing a total of $m_{si}$ tokens
and accompanied by sentiment $y_{si}\in \{-1,0,1\}$, corresponding to
negative, neutral, and positive sentiment respectively.  Collapsed
counts for each politician-sentiment combination are obtained as
$x_{syj} = \sum_{i: y_{si} = y} x_{sij}$ for each token $j$.  This
yields 17 `observations': each of three sentiments for five
politicians, plus positive and negative emoticon tweets.  The
multinomial inverse regression model for sentiment-$y$ text counts
directed towards politician $s$ is then $\bm{x}_{sy} \sim
\mr{MN}(\bm{q}_{sy}, m_{sy})$, $q_{syj} = e^{\eta_{scy}}/\sum_{l=1}^p
e^{\eta_{syl}}$ for $j=1\ldots p$, with linear equation
\begin{equation}\label{pols-mnir}
 \eta_{syj} = \alpha_{0j} + \alpha_{sj} +
y(\varphi_{0j} + \varphi_{sj}).
\end{equation}
Politician-specific terms are set to zero for emoticon tweets (which
are not associated with a specific politician), say $s=e$, such that
$\eta_{eyj} = \alpha_{0j} + y\varphi_{0j}$ as a generic sentiment
model.  Thus all text is centered on  main effects in
$\bs{\alpha}_{0}$ and $\bs{\varphi}_0$, while interaction terms
 $\bs{\alpha}_s$ and  $\bs{\varphi}_s$ are
identified only through their corresponding turk-scored political
sentiment sample.

Results in \cite{Tadd2012a} show that $\bm{x}'[\bs{\varphi}_{0},
\bs{\varphi}_{s}]$ is sufficient for sentiment when $\bm{x}$ is drawn
from the collapsed count model implied by (\ref{pols-mnir}).  Thus
following the same logic behind our univariate SR scores in
(\ref{basic-sr}), $\bm{z}_{i} = [\bm{z}_{i0},\bm{z}_{is}] =
\bm{f}_{i}'[\bs{\varphi}_{0}, \bs{\varphi}_{s}]$ is a bivariate
sufficient reduction score for tweet $i$ on politician $s$.  The
forward   model is again proportional-odds
logistic regression,
\begin{equation}\label{pols-fwd}
\mr{p}( y_{i} \leq c) = 1/(1 + \exp[ \beta_0 z_{i0} +
\beta_s z_{is} - \gamma_c ]),
\end{equation}
with main $\beta_0$ and subject $\beta_s$ effects.  Note the absence
of subject-specific $\gamma_{sc}$: a tweet containing no significant
tokens (such that $z_{i0} = z_{is} = 0$) is assigned probabilities
according to the overall aggregation of tweets.  Such `empty' tweets
have $\mr{p}(-1) = 0.25$, $\mr{p}(0) = 0.65$, and $\mr{p}(1) = 0.1$ in
the fitted model of Section \ref{analysis}, and are thus all
classified as `neutral'.

\subsection{Notes on MNIR estimation}

Estimation of MNIR models like those in (\ref{basic-mnir}) and
(\ref{pols-mnir}) follows exactly the procedures of \cite{Tadd2012a},
and the interested reader should look there for detail.  Briefly, we
apply the {\it gamma lasso} estimation algorithm, which corresponds to
MAP estimation under a hierarchical gamma-Laplace coefficient prior
scheme. Thus, and this is especially important for the interaction
models of Section \ref{mnir}.1, parameters are estimated as exactly
zero until a large amount of evidence has accumulated.  Optimization
proceeds through coordinate descent and, along with the obvious
efficiency derived from collapsing observations, allows for estimation
of single-factor SR models with hundreds of thousands of tokens in
mere seconds.  The more complicated interaction model in
(\ref{pols-mnir}) can be estimated in less than 10 minutes.

To restate the MNIR strategy, we are using a simple but very
high-dimensional (collapsed count) model to obtain a useful but
imperfect text summary for application in low dimensional sentiment
regression. MNIR works because the multinomial is a useful
representation for token counts, and this model assumption increases
efficiency by introducing a large amount of information 
about the functional relationship between text and sentiment into the
prediction problem. Implicit here is an assumption that ad-hoc
forward regression  can compensate for mis-application of
population-average summary projections to individual document counts.
\cite{Tadd2012a} presents  empirical evidence that this holds
true in practice, with MNIR yielding higher quality prediction at
lower computational cost when compared to a variety of text regression
techniques. However the design algorithms of this article are not
specific to MNIR and can be combined with any sentiment prediction
routine.

\section{Topic-optimal design}
\label{design}

Recall the introduction's pool-based design problem: choosing from the
full sample of 2.1 million political tweets a subset to be scored, on
mechanical turk, as either negative, neutral, or positive about the
relevant politician.

A short review of some relevant literature on active learning and
experimental design is in the appendix.  In our specific situation of a
very high dimensional input space (i.e a large vocabulary), effective
experimental design is tough to implement.  Space-filling is impractical
since limited sampling will always leave a large distance between
observations.  Boundary selection -- where documents with roughly
equal sentiment-class probabilities are selected for scoring -- leads
to samples that are very sensitive to model fit and is impossible in
early sampling where the meaning of most terms is unknown (such that
the vast majority of documents lie on this boundary).  Moreover,
one-at-a-time point selection implies sequential algorithms that
scale poorly for large applications, while more elaborate active
learning routines which solve for optimal batches of new points tend
to have their own computational limits in high dimension.  Finally,
parameter and predictive uncertainty -- which are relied upon in many
active learning routines -- is difficult to quantify in complicated
text regression models; this includes MNIR, in which the posterior is
non-smooth  and is accompanied by an ad-hoc forward
regression step.  The vocabulary is also growing with sample size and
a full accounting of uncertainty about sentiment in unscored texts
would depend heavily on a prior model for the meaning of previously
unobserved words.

While the above issues make tweet selection difficult, we do have an
advantage that can be leveraged in application: a huge pool of
unscored documents.  Our solution for text sampling is thus to look at
space-filling or optimal design criteria (e.g., D-optimality) but on a
reduced dimension factor decomposition of the covariate space rather
than on $\bm{X}$ itself.  That is, although the main goal is to learn $\bs{\Phi}$ 
for the sentiment projections of Section \ref{mnir},  this cannot be done until 
enough documents are scored and we instead look to space-fill on an
{\it unsupervised} factor structure that can be estimated without labelled examples.
This leads to to what we call {\it
  factor-optimal design}.  Examples of this approach include
\citet{GalvMaccBezz2007} and \citet{ZhanEdga2008}, who apply optimal
design criteria on principal components, and \citet{DavyLuz2007}, a
text classification contribution that applies active learning criteria
to principal components fit for word counts.  The proposal here is to
replace generic principal component analysis with text-appropriate
topic model factorization.

\subsection{Multinomial topic factors}

A $K$-topic model \citep{BleiNgJord2003} represents each vector of
document token counts, $\bm{x}_i \in \{\bm{x}_{1}\ldots \bm{x}_n\}$
with total $m_i = \sum_{j=1}^p x_{ij}$, as a multinomial factor
decomposition
\begin{equation}\label{eq:tpc}
\bm{x}_i \sim \mr{MN}(\omega_{i1} \bs{\theta}_{1} + \ldots + \omega_{iK}
\bs{\theta}_{K}, m_i)
\end{equation}
where topics $\bs{\theta}_k = [\theta_{k1} \cdots \theta_{kp}]'$ and
weights $\bs{\omega}_i$ are probability vectors.  Hence, each topic
$\bs{\theta}_k$ -- a vector of probabilities over words or phrases --
corresponds to factor `loadings' or `rotations' in the usual factor
model literature.  Documents are thus characterized through a
mixed-membership weighting of topic factors and $\bs{\omega}_i$ is a
reduced dimension summary for $\bm{x}_i$.

Briefly, this approach assumes independent prior
distributions for each probability vector,
\begin{equation}\label{eq:prior}
\bs{\omega}_i \stackrel{iid}{\sim} \mr{Dir}(1/K),~i=1\ldots n,~~\text{and}~~
\bs{\theta}_k  \stackrel{iid}{\sim}
\mr{Dir}(1/(Kp)),~k=1\ldots K,
\end{equation}
where $\bs{\theta} \sim \mr{Dir}(\alpha)$ indicates a Dirichlet
distribution with concentration parameter $\alpha$ and density
proportional to $\prod_{j=1}^{\mr{dim}(\bs{\theta})}
\theta_j^\alpha$. These $\alpha < 1$ specifications encourage a few
dominant categories among mostly tiny probabilities by placing
weight at the edges of the simplex.  The particular specification in
(\ref{eq:prior}) is chosen so that prior weight, measured as the sum
of concentration parameters multiplied by the dimension of their
respective Dirichlet distribution, is constant in both $K$ and $p$
(although not in $n$).  The model is estimated through  posterior
maximization as in \citet{Tadd2012b}, and we employ a Laplace
approximation for simulation from the conditional posterior for
$\bs{\Omega}$ given $\bs{\Theta} = [\bs{\theta}_1 \cdots
\bs{\theta}_K]$.  The same posterior approximation allows us to
estimate Bayes factors for potential values of $K$, and we use this to
{\it infer} the number of topics from the data.  Details are in
Appendix \ref{bayes}.

\subsection{Topic D-optimal design}

As a general practice, one can look to implement any space filling
design in the $K$ dimensional $\bs{\omega}$-space.  For the current
study, we focus on D-optimal design rules that seek to maximize the
determinant of the information matrix for linear regression; the
result is thus loosely optimal under the assumption that sentiment has
a linear trend in this representative factor space.  The algorithm
tends to select observations that are at the edges of the topic space.
An alternative option that may be more robust to sentiment-topic
nonlinearity is to use a latin hypercube design; this will lead to a
sample that is spread evenly throughout the topic space.

In detail, we
seek to select a design of documents $\{i_1 \ldots i_T\} \subset
\{1\ldots n\}$ to maximize the topic information determinant $D_T =
|\bs{\Omega}_T'\bs{\Omega}_T|$, where $\bs{\Omega}_T = [\bs{\omega}_1
\cdots \bs{\omega}_T]'$ and $\bs{\omega}_t$ are topic weights
associated with document $i_t$.  Since construction of exact D-optimal
designs is difficult and the algorithms are generally slow
\citep[see][for an overview of both exact and approximate optimal
design]{AtkiDone1992}, we use a simple greedy search to obtain an {\it
  ordered} list of documents for evaluation in a near-optimal design.

Given $D_t = |\bs{\Omega}_t'\bs{\Omega}_t|$ for a current sample of
size $t$, the topic information determinant after adding $i_{t+1}$ as
an additional observation is
\begin{equation}\label{dup}
D_{t+1} = \left|\bs{\Omega}_t'\bs{\Omega}_t + \bs{\omega}_{t+1}'
\bs{\omega}_{t+1}\right|  = D_{t}\left( 1 +
  \bs{\omega}_{t+1}' \left(\bs{\Omega}_t'\bs{\Omega}_t\right)^{-1} \bs{\omega}_{t+1}\right),
\end{equation}
due to a standard linear algebra identity.  This implies that,  given
$\bs{\Omega}_t$ as the topic matrix for your currently evaluated
documents, $D_{t+1}$ is maximized simply by choosing $i_{t+1}$ such that
\begin{equation}\label{max}
\bs{\omega}_{t+1}  = \mr{argmax}_{\{\bs{\omega}\in
  \bs{\Omega}/\bs{\Omega}_t\}} ~~\bs{\omega}' \left(\bs{\Omega}_t'\bs{\Omega}_t\right)^{-1} \!\!\bs{\omega}
\end{equation}
Since the topic weights are a low ($K$) dimensional summary, the
necessary inversion $\left(\bs{\Omega}_t'\bs{\Omega}_t\right)^{-1}$ is
on a small $K\times K$ matrix and will not strain computing resources.
This inverted matrix provides an operator that can quickly be applied
to the pool of candidate documents (in parallel if desired), yielding
a simple score for each that represents the proportion by which its
inclusion increases our information determinant.

For the recursive equation in (\ref{max}) to apply, the design must be
initially seeded with at least $K$ documents, such that
$\bs{\Omega}_t'\bs{\Omega}_t$ will be non-singular.  We do this by
starting from a simple random sample of the first $t=K$ documents
(alternatively, one could use more principled space-filling in factor
space, such as a latin hypercube sample).  Note
that again topic-model dimension reduction is crucial: for our greedy
algorithm to work in the full $p$ dimensional token space, we would
need to sample $p$ documents before having an invertible information
matrix.  Since this would typically be a larger number of documents
than desired for the full sample, such an approach would never move
beyond the seeding stage.

In execution of this design algorithm, the topic weights for each
document must be estimated.  In what we label MAP topic D-optimal
design, each $\bs{\omega}_i$ for document $i$ is fixed at its MAP
estimate as described in Section \ref{design}.1.  As an alternative,
we also consider a {\it marginal} topic D-optimality wherein a set of
topic weights $\{\bs{\omega}_{i1}\ldots \bs{\omega}_{iB}\}$ are
sampled for each document from the approximate posterior in Appendix
A.1, such that recursively D-optimal documents are chosen to
maximize the {\it average} determinant multiplier over this set.
Thus, instead of (\ref{max}), marginal D-optimal $i_{t+1}$ is
selected to maximize $\frac{1}{B}\sum_b \bs{\omega}_{i_{t+1}b}'
\left(\bs{\Omega}_t'\bs{\Omega}_t\right)^{-1}
\!\!\bs{\omega}_{i_{t+1}b}$.

\subsection{Note on the domain of factorization}

The basic theme of this design framework is straightforward: fit an
unsupervised factor model for $\bm{X}$ and use an optimal design rule
in the resulting factor space.  Given a single sentiment
variable, as in examples of Section \ref{examples}, the $\bm{X}$ to
be factorized is simply the entire text corpus.

Our political twitter case study introduces the added variable of
`politician', and it is no longer clear that a single shared
factorization of all tweets is appropriate.  Indeed, the interaction
model of Section \ref{mnir}.2 includes parameters (the $\alpha_{sj}$
and $\varphi_{sj}$) that are only identified by tweets on the
corresponding politician.  Given the massive amount of existing data
from emoticon tweets on the other model parameters, any
parameter learning from new sampling will be concentrated on these
interaction parameters.  Our solution in Section \ref{analysis} is to
apply stratified sampling: fit independent factorizations to each
politician-specific sub-sample of tweets, and obtain D-optimal designs
on each.  Thus we ensure a scored sample of a chosen size for each
individual politician.

\section{Example Experiment}
\label{examples}

To illustrate this design approach, we consider two simple
text-sentiment examples.  Both are detailed in
\cite{Tadd2012a,Tadd2012b}, and available in the {\cd textir} package
for {\cd R}. {\it Congress109} contains 529 legislators' usage counts
for each of 1000 phrases in the $109^{th}$ US Congress, and we
consider party membership as the `sentiment' of interest: $y=1$ for
Republicans and $0$ otherwise (two independents caucused with
Democrats).  {\it We8there} consists of counts for 2804 bigrams in
6175 online restaurant reviews, accompanied by restaurant {\it
  overall} rating on a scale of one to five.  To mimic the motivating
application, we group review sentiment as negative ($y=-1$) for
ratings of 1-2, neutral ($y=0$) for 3-4, and positive ($y=1$) for 5
(average rating is 3.95, and the full 5-class analysis is in
\citealt{Tadd2012a}).  Sentiment prediction follows the single-factor MNIR
procedure of Section \ref{mnir}, with binary logistic forward
regression $\ds{E}[y_i] = \exp[ \gamma + \beta z_i]/(1 + \exp[ \gamma
+ \beta z_i] )$ for the congress data, and proportional-odds logistic
regression $\mr{p}(y_i \leq c) = \exp[ \gamma_c - \beta z_i]/(1 +
\exp[ \gamma_c - \beta z_i] )$, $c=-1,0,1$ for the we8there data.

We fit $K =$ 12 and 20 topics respectively to the congress109 and
we8there document sets.  In each case, the number of topics is chosen
to maximize the approximate marginal data likelihood, as detailed in
the appendix and in \cite{Tadd2012b}.  Ordered sample designs were
then selected following the algorithms of Section \ref{design}.2: for
MAP D-optimal, using MAP topic weight estimates, and for marginal
D-optimal, based upon approximate posterior samples of 50 topic
weights for each document.  We also consider principal component
D-optimal designs, built following the same algorithm but with topic
weights replaced by the same number (12 or 20) of principal components
directions fit on token frequencies $\bm{f}_i = \bm{x}_i/m_i$.
Finally, simple random sampling is included as a baseline, and was
used to seed each D-optimal algorithm with its first $K$ observations.
Each random design algorithm was repeated 100 times.

\begin{figure}[t]
\hskip -.4cm\includegraphics[width=6.6in]{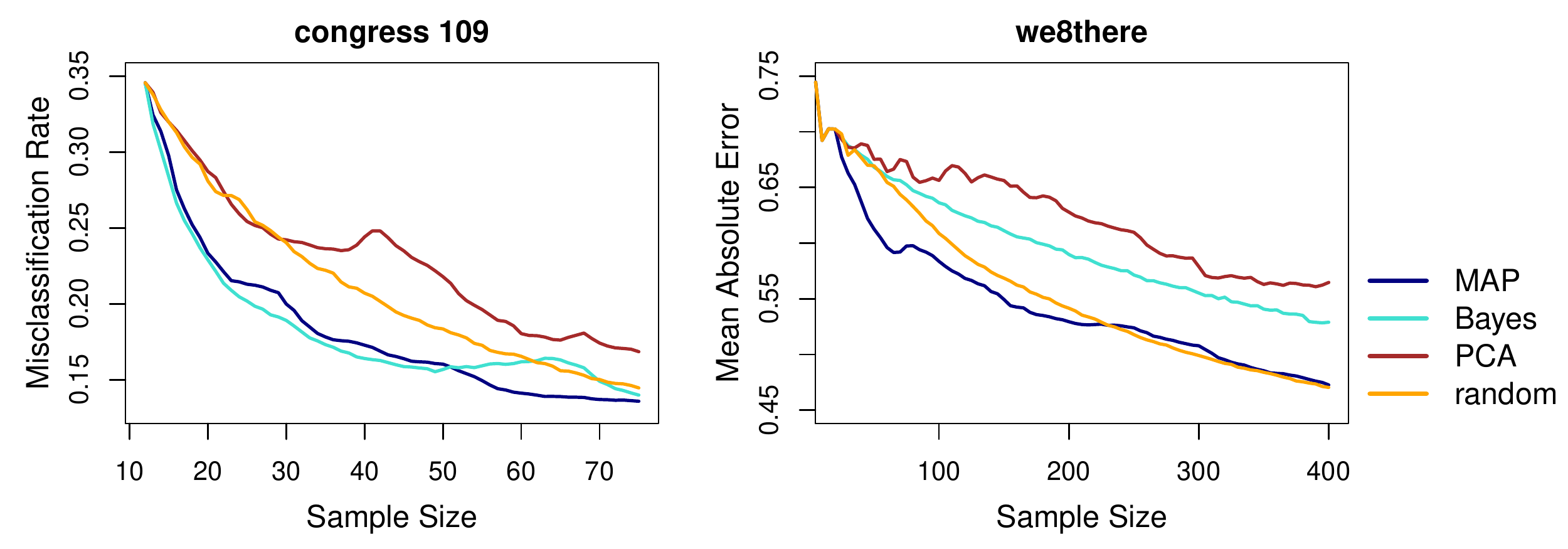}
\caption{\label{experiment} Average error rates on 100 repeated
  designs for the 109$^{th}$ congress and we8there examples.
  `MAP' is D-optimal search on MAP estimated topics; `Bayes' is our
  search for marginal D-optimality when sampling from the topic
  posterior; `PCA' is the same D-optimal search in principal
  components factor space; and `random' is simple random
  sampling. Errors are evaluated over the entire dataset.}
\end{figure}

Results are shown in Figure \ref{experiment}, with average error rates
(misclassification for congress109 and mean absolute error for
we8there) reported for maximum probability classification over the
entire data set.  The MAP D-optimal designs perform better than simple
random sampling, in the sense that they provide faster reduction in
error rates with increasing sample size.  The biggest improvements are
in early sampling and error rates converge as we train on a larger
proportion of the data.  There is no advantage gained from using a
principal component (rather than topic) D-optimal design, illustrating
that misspecification of factor models can impair or eliminate their
usefulness in dimension reduction.  Furthermore, we were surprised to
find that, in contrast with some previous studies on active learning
\citep[e.g.][]{TaddGramPols2011}, averaging over posterior uncertainty
did not improve performance: the MAP D-optimal design does as well or
better than the marginal alternative, which is even outperformed by
random sampling in the we8there example. Our hypothesis is that, since
conditioning on $\bs{\Theta}$ removes dependence across documents,
sampling introduces Monte Carlo variance without providing any
beneficial information about correlation in posterior uncertainty.
Certainly, given that the marginal algorithm is also much more time
consuming (with every operation executed $B$ times in addition to the
basic cost of sampling), it seems reasonable to focus on the MAP
algorithm in application.

\section{Analysis of Political Sentiment in Tweets}
\label{analysis}

This section describes selection of tweets for sentiment scoring from
the political Twitter data described in Section \ref{data}, under the
design principles outlined above, along with an MNIR analysis of the
results and sentiment prediction over the full collection.

\subsection{Topic factorization and D-optimal design}

As the first step in experimental design, we apply the topic
factorization of Section \ref{design}.1 independently to each
politician's tweet set.  Using the Bayes factor approach of
\cite{Tadd2012b}, we tested $K$ of 10, 20, 30 and 40 for each
collection and, in every case, selected the simple $K=10$ model as
most probable.  Although this is a smaller topic model than often seen
in the literature, we have found that posterior evidence tends to
favor such simple models in corpora with short documents
\citep[see][for discussion of information increase with
$m_i$]{Tadd2012b}.

Across politicians, the most heavily used topic (accounting for about
20\% of words in each case) always had {\cd com}, {\cd
  http}, and {\cd via} among the top five tokens by
topic lift -- the probability of a token within a topic over its
overall usage proportion.  Hence, these topics appear to represent a
Twitter-specific list of stopwords.  The other topics are a mix of
opinion, news, or user specific language.  For example, in the Gingrich
factorization one topic accounting for 8\% of text with top tokens {\cd
  herman}, {\cd cain}, and {\cd endors} is focused on Herman Cain's
endorsement, {\cd \#teaparty} is a top token in an 8\% topic that
appears to contain language used by self identified members of the Tea
Party movement (this term loads heavily in a single topic for each
politician we tracked), while another topic with {\cd @danecook} as
the top term accounts for 10\% of traffic and is dominated by posts of
unfavorable jokes and links about Gingrich by the comedian Dane Cook
(and forwards, or `retweets', of these jokes by his followers).

Viewing the sentiment collection problem through these interpreted
topics can be useful: since a D-optimal design looks (roughly) for
large variance in topic weights, it can be seen as favoring tweets
on single topics (e.g., the Cain endorsement) or rare
combinations of topics (e.g., a Tea Partier retweeting a Dane Cook
joke).  As a large proportion of our data are retweets (near 40\%),
scoring those sourced from a single influential poster can yield a
large reduction in predictive variance, and tweets containing
contradictory topics help resolve the relative weighting of words.
In the end, however, it is good to remember that the topics do not
correspond to subjects in the common understanding, but are simply
loadings in a multinomial factor model.  The experimental design described
in the next section treats the fitted topics as such.

\subsection{Experimental  design and sentiment collection}

Using the MAP topic D-optimal algorithm of Section \ref{design}.2,
applied to each politician's topic factorization,
we built ordered lists of tweets to be scored on Mechanical Turk: 500
for each Republican primary candidate, and 750 for Obama.  Worker
agreement rates varied from 78\% for Obama to 85\% for Paul, leading
to sample sizes of 406 for Romney, 409 for Santorum, 418 for Gingrich,
423 for Paul, and 583 for Obama.

Unlike  the experiment of Section \ref{examples}, we have no ground truth
for evaluating model performance across samples without having to pay
for a large amount of turk scoring.  Instead, we propose two metrics:
the number of non-zero politician specific loadings $\varphi_{js}$,
and the average entropy $-\sum_{c=-1,0,1} \mr{p}_{c} \log(\mr{p}_{c})$
across tweets for each politician, where $\mr{p}_{c} = \mr{p}(y =c)$
is based on the forward proportional-odds regression described below
in \ref{analysis}.2.  We prefer the former for measuring the {\it
  amount of sample evidence} -- the number of tokens estimated as
significant for politician-specific sentiment in gamma-lasso penalized
estimation -- as a standard statistical goal in design of experiments,
but the latter corresponds to the more common machine learning metric
of classification precision (indeed, entropy calculations inform many
of the close-to-boundary active learning criteria in Appendix \ref{back}).

\begin{figure}[t]
\includegraphics[width=6.4in]{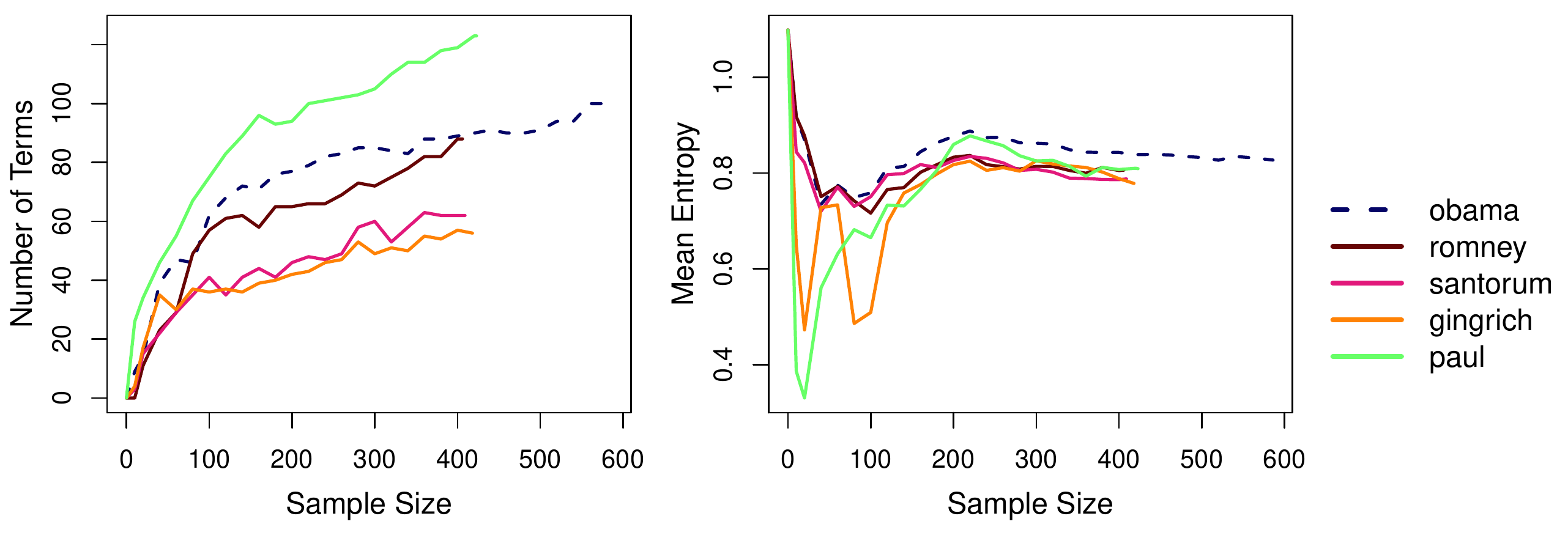}
\caption{\label{learning} Learning under the MAP topic D-optimal
  design.  For increasing numbers of scored tweets added from the
  ranked design, the left shows the number of significant (nonzero) loadings in the direction of
  politician-specific sentiment and the right shows mean entropy $-\sum
  \mr{p}_c \log(\mr{p}_{c})$ over the full sample.   As in
  Figure \ref{volume}, blue is Obama, orange Romney, red Santorum, pink
  Gingrich, and green Paul.}
\end{figure}

Results are shown in Figure \ref{learning} for the sequential addition
of scored tweets from the design-ranked Turk results (sentiment
regression results are deferred until Section \ref{analysis}.3).  On the left, we
see that there is a steady climb in the number of nonzero
politician-specific loadings as the sample sizes increase.  Although
the curves flatten with more sampling, it does appear that had we
continued spending money on sending tweets to the Turk it would have
led to larger politician-sentiment dictionaries.  The right plot shows
a familiar pattern of early overfit (i.e., underestimated
classification variance) before the mean entropy begins a slower
steady decline from $t=200$ onwards.

\subsection{MNIR for subject-specific sentiment analysis}

After all Turk results are incorporated, we are left with 2242 scored
political tweets, plus the 1.6 million emoticon tweets, and a 5566
token vocabulary.  This data were used to fit the
politician-interaction MNIR model detailed in Section \ref{mnir}.2.

The top ten politician-specific loadings
($\varphi_{sj}$) by absolute value are shown in Table \ref{loadings}
(recall that these are the effect on log odds for a unit increase in
sentiment; thus, e.g., negatively loaded terms occur more frequently
in negative tweets).  This small sample shows some large
coefficients, corresponding to indicators for users or groups, news
sources and events, and various other labels.  For example, the Obama
column results suggest that his detractors prefer to use `GOP' as
shorthand for the republican party, while his supporters simply use
`republican'. However, one should be cautious about interpretation:
these coefficients correspond to the partial effects of sentiment on the
usage proportion for a term {\it given} corresponding change in
relative frequency for all other terms. Moreover, these are only
estimates of average correlation; this analysis is not
intended to provide a causal or long-term text-sentiment model.

\begin{figure}[t]
\begin{minipage}{3.2in}
\includegraphics[width=3.2in]{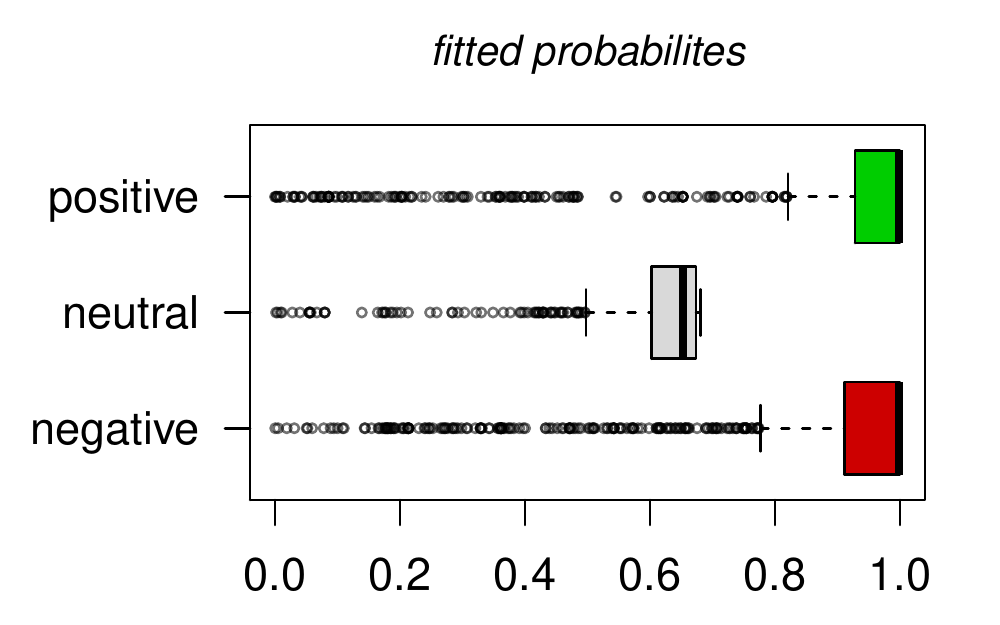}
\caption{\label{fit} In-sample sentiment fit: the forward
  model probabilities for each observation's true  category. }
\end{minipage}
~~~~~
\begin{minipage}{2.8in}\small
\vspace{1cm}
\begin{tabular}{lccc}
\multicolumn{1}{c}{} & $\mr{cor}(z_s,z_0)$ & $\mr{sd}(z_s)$ & $\bar z_s$ \\
\\[-1.5ex]\cline{2-4}\\[-1.5ex]
\sf obama&-0.12&0.31&0.1\\
\sf romney&0.17&0.23&-0.07\\
\sf santorum&0.16&0.19&-0.19\\
\sf gingrich&-0.07&0.26&-0.06\\
\sf paul&0.07&0.16&0.1\\
\sf emoticons &---&0.06&0.006
\end{tabular}
\vskip .25cm
\captionof{table}{\label{zsmry} {\it Full} sample summary statistics for
  politician-specific sufficient reduction scores.  }
\end{minipage}
\end{figure}

Summary statistics for fitted SR scores are shown in Table
\ref{zsmry}.  Although we are not strictly forcing orthogonality on
the factor directions -- $z_{0}$ and $z_s$, say the emotional and
political sentiment directions respectively -- the political scores
have only weak correlation (absolute value $<$ 0.2) with the generic
emotional scores.  This is due to an MNIR setup that estimates
politician-specific loadings $\varphi_{sj}$ as the sentiment effect on
language about a given politician {\it after} controlling for generic
sentiment effects.  Notice that there is greater variance in political
scores than in emotional scores; this is due to a few large token
loadings that arise by identifying particular tweets (that are heavily
retweeted) or users that are strongly associated with positive or
negative sentiment.  However, since we have far fewer scored political
tweets than there are emoticon tweets, fewer token-loadings are
non-zero in the politician-specific directions than in the generic
direction: $\bs{\varphi}_0$ is only 7\% sparse, while the
$\bs{\varphi}_s$ are an average of 97\% sparse.

\begin{table}[t]
\vspace{.1cm}
\centering\small
\begin{tabular}{cl|cl|cl|cl|cl}
  \multicolumn{2}{c}{\normalsize \sc Obama}
  &\multicolumn{2}{c}{\normalsize \sc Romney}
  &\multicolumn{2}{c}{\normalsize \sc Santorum}
  &\multicolumn{2}{c}{\normalsize \sc Gingrich}
  &\multicolumn{2}{c}{\normalsize \sc Paul}\\
[-1.5ex]\\
\cd republican&15.5&\cd fu&-10&\cd @addthi&-11.5&\cd bold&10.6&\cd \#p2&11.1\\
 \cd gop&-13.2&\cd 100\%&-9.6&\cd @newtgingrich&-9.9&\cd mash&-10&\cd \#teaparti&11\\
 \cd \#teaparti&-12.7&\cd lover&-9.4&\cd clown&-9.4&\cd ap&9.9&\cd ht&10\\
 \cd \#tlot&-11.9&\cd quot&-9.4&\cd @youtub&-9.2&\cd obama&9.9&\cd airplan&9.6\\
 \cd economi&11&\cd anytim&-9.2&\cd polit&-8.7&\cd campaign&-9.9&\cd legal&-9.5\\
 \cd cancer&10&\cd abt&-8.6&\cd speech&-8.6&\cd lesbian&-9.7&\cd paypal&7.4\\
 \cd cure&9.6&\cd lip&-8.5&\cd opportun&-8.2&\cd pre&9.5&\cd flight&6.9\\
 \cd ignor&9.2&\cd incom&-8.4&\cd disgust&-8.2&\cd bid&9.5&\cd rep&6.7\\
 \cd wors&9.2&\cd januari&8.1&\cd threw&-7.4&\cd recip&-9.2&\cd everyth&-6.4\\
 \cd campaign&9.2&\cd edg&8&\cd cultur&-7.3&\cd america&9.1&\cd debat&6
\end{tabular}
\caption{ Top ten politician-specific token loadings $\varphi_{sj}$ by
  their absolute value in
  MNIR. \label{loadings}}
\end{table}

Figure \ref{fit} shows fitted values in forward proportional-odds
logistic regression for these SR scores.  We observe some very high
fitted probabilities for both true positive and negative tweets,
indicating again that the analysis is able to identify a subset of
similar tweets with easy sentiment classification.  Tweet
categorization as neutral corresponds to an absence of evidence in
either direction, and neutral tweets have fitted $\mr{p}(0)$ with mean
around 0.6.  In other applications, we have found that a large number
of `junk' tweets (e.g., selling an unrelated product) requires
non-proportional-odds modeling to obtain high fitted neutral
probabilities, but there appears to be little junk in the current
sample.  As an aside, we have experimented with adding `junk' as a
fourth possible categorization on Mechanical Turk, but have been
unable to find a presentation that avoids workers consistently getting
confused between this and `neutral'.

\begin{table}[t]
\vspace{.2cm}
\small\hspace{-.3cm}
\begin{tabular}{lccccccccc}
  &\multicolumn{2}{c}{\normalsize \sc Intercepts $\gamma_c$} &
  &\multicolumn{6}{c}{\normalsize \sc SR score coefficients $\beta_0$,
    $\beta_s$} \\
\cline{2-3} \cline{5-10} \\[-2ex]
  &  $\leq -1$ & \hspace{-.3cm}$\leq 0$ && \cd emoticons & \cd obama & \cd romney & \cd
  santorum & \cd gingrich & paul\\ [-2ex]\\
  Estimate &  -1.1 {\smaller (0.1)}&\hspace{-.3cm} 2.2 {\smaller  (0.1)}&&8.3 {\smaller  (1.1)}&4.9 {\smaller  (0.5)}&
  5.6 {\smaller  (0.5)}&5.8 {\smaller  (0.5)}&7.9 {\smaller
    (1.0)}&11.9 {\smaller  (1.1)}\\ [-2.5ex]\\
$\beta \times \bar z_s$ & & & & 0.0 & 0.5 & -0.4& -1.1& -0.5 &  1.2 \\
{$\exp[\beta \times \mr{sd}(z)]$}\hspace{-.2cm} & & & & 1.6 & 4.5 & 3.6 & 2.9 & 7.7 & 6.4
\end{tabular}
\caption{ MAP estimated parameters and the conditional standard
  deviation (ignoring variability in $\bm{z}$) in the forward proportional-odds
  logistic regression $\mr{p}( y_{i} \leq c) = (1 + \exp[ \beta_0 z_{i0} +
  \beta_s z_{is} - \gamma_c])^{-1}$, followed by the average effect on
  log-odds for each sufficient reduction score and exponentiated
  coefficients scaled according to the corresponding full-sample score
  standard deviation.   \label{coef}}
\end{table}

The forward parameters are MAP estimated, using the {\cd arm} package
for {\cd R} \citep{GelmSuYajiHillPittKermZhen2012}, under diffuse
$t$-distribution priors; these estimates are printed in Table
\ref{coef}, along with some summary statistics for the implied effect
on the odds of a tweet being at or above any given sentiment level.
The middle row of this table contains the average effect on log-odds
for each sufficient reduction score: for example, we see that Santorum
tweet log-odds drop by an average of -1.1 ($e^{-1.1} \approx 0.3$)
when you include his politician-specific tweet information.  The
bottom row shows implied effect on sentiment odds scaled for a
standard deviation increase in each SR score direction: an extra
deviation in emotional $z_0$ multiplies the odds by $e^{0.5} \approx
1.6$, while a standard deviation increase in political SR scores
implies more dramatic odds multipliers of 3 (Santorum) to 8
(Gingrich).  This agrees with the fitted probabilities of Figure
\ref{fit}, and again indicates that political directions are
identifying particular users or labels, and not `subjective language'
in the general sense.

\begin{figure}[t]
\includegraphics[width=6.3in]{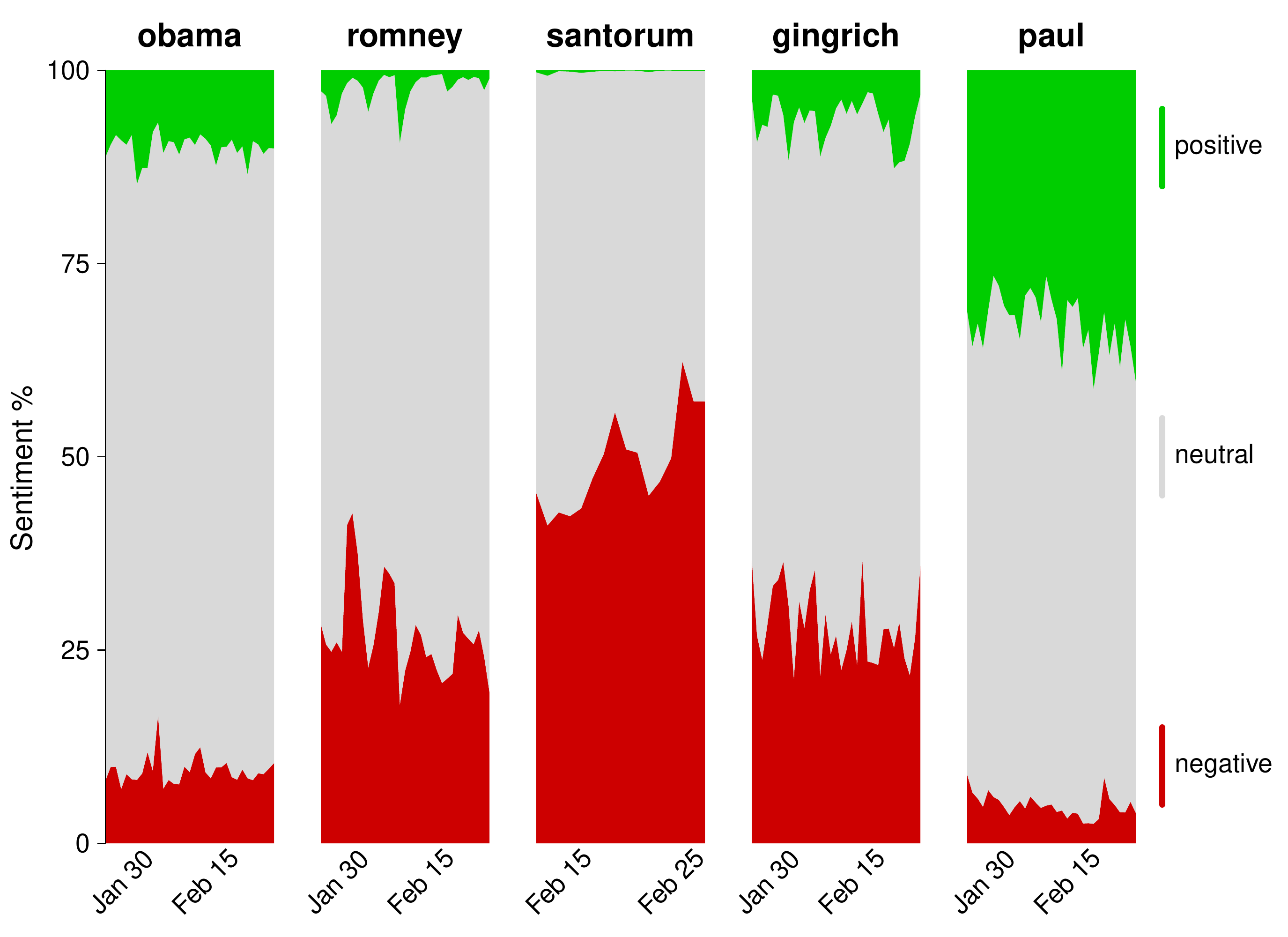}
\caption{\label{sentiment} Twitter sentiment regression full-sample
  predictions.  Daily tweet count percentages by sentiment
  classification are shown with green for positive, grey neutral, and
  red  negative.  }
\end{figure}

Figure \ref{sentiment} shows predicted sentiment classification for
each of our 2.1 million collected political tweets, aggregated by day
for each politician-subject. In each case, the majority of traffic
lacks enough evidence in either direction, and is classified as
neutral.  However, some clear patterns do arise.  The three
`mainstream' Republicans (Romney, Santorum, Gingrich) have far more
negative than positive tweets, with Rick Santorum performing worst.
Libertarian Ron Paul appears to be relatively popular on Twitter,
while President Obama is the only other politician to receive
(slightly) more positive than negative traffic.  It is also possible
to match sentiment classification changes to events; for example,
Santorum's negative spike around Feb 20 comes after a weekend of new
agressive speeches in which he referenced Obama's `phony theology' and
compared schools to `factories', among other lines that generated
controversy.

Finally, note for comparison that without the interaction terms (i.e,
with only {\it score} as a covariate in inverse regression), the
resulting univariate SR projection is dominated by emoticon-scored
text.  These projections turn out to be a poor summary of sentiment in
the political tweets: there is little discrimination between SR scores
across sentiment classes, and the in-sample mis-classification rate jumps
to 42\% (from 13\% for the model that uses politician-specific
intercepts).  Fitted class probabilities are little different from
overall class proportions, and with true neutral tweets being less
common (at 22\% of our turk-scored sample) the result is that all
future tweets are unrealistically predicted as either positive or
negative.

\section{Discussion}
\label{discussion}

This article makes two simple proposals for text-sentiment analysis.
First, looking to optimal design in topic factor space can be useful
for choosing documents to be scored.  Second, sentiment can be
interacted with indicator variables in MNIR to allow subject-specific
inference to complement information sharing across generic
sentiment.

Both techniques deserve some caution.  Topic D-optimal design ignores
document length, even though longer documents can be more informative;
this is not a problem for the standardized Twitter format, and did not
appear to harm design for our illustrative examples, but it could be
an issue in other settings.  In the MNIR analysis, we have observed that
subject-specific sentiment loadings (driven in estimation by small
sample subsets) can be dominated by news or authors specific to the
given sample. While this is not technically overfit, since it is
finding persistent signals in the current time period, it indicates
that one should constantly update models when using these techniques
for longer-term prediction.

A general lesson from this study is that traditional statistical techniques, such
as experimental  design and variable interaction, will apply in new
areas like text mining when used in conjunction with careful
dimension reduction.  Basic statistics principles can then be relied
upon  to build optimal predictive models and to assess their risk
and sensitivity in application.

\appendix

\section{Appendix}

\subsection{Review: active learning and optimal
  design}
\label{back}

The literature on text sampling is focused on the type of
design of experiments that is referred to as {\it active learning} in
machine learning.  There are two main components: optimality and
adaptation.  In the first, new input locations are chosen with regard
to the functional form of the regression model (and possibly
current parameter fits) and, in the second,
data are added sequentially wherever it is most needed
according to a specific design criterion.  Early examples of this
framework include the contributions of \citet{MacK1992}, sampling
always the new point with highest predictive response variance, and
\citet{Cohn1996}, choosing new inputs to maximize the expected
reduction in predictive variance.

In text analysis, the work of \cite{TongKoll2001} on active learning for text
classification with support vector machines has been very
influential.  Here, the next evaluated point should be that which
minimizes the expected {\it version space} -- the set of classification
rules which imply perfect separation on the current sample (standard
for support vector machines, kernel expansions of the covariate space ensure
such separation is possible).  Hence, the criterion is analogous to
Cohn's expected predictive variance, but for an overspecified
algorithm without modeled variance.  Tong and Koller propose three
ways to find an approximately maximizing point, the most practical of
which is labelled {\cd simple}: choose the point closest to the
separating hyperplane.  This is equivalent to the algorithm of
\citet{SchoCohn2000}.

In general, algorithms within the large literature on active learning for text
regression, and similar classification problems (e.g. image sorting),
follow the same theme: define a metric that summarizes `response variability'
for your given prediction technique, and sequentially sample inputs
which maximize this metric or its expected reduction over some
pre-defined set.  For example, \cite{Bish2009} minimize approximate
expected classification loss in an algorithm nearly equivalent to {\cd
  simple}, \citet{LierTade1997} generate predictions from a
`committee' of classifiers and sample points where there is
disagreement about the class label, and \citet{HoluPeroBurl2008}
sample to maximize the reduction in expected entropy.  Since all of
these methods seek to choose points near the classification boundary,
it is often desirable to augment the active learning with points from a
space filling design \citep[e.g.][]{Hu2010}.  Under fully Bayesian
classifier active learning, as in \cite{TaddGramPols2011},
such `exploration' is automatic through accounting for posterior
uncertainty about class probabilities.

A related literature from statistics is that on {\it optimal design},
wherein sampling is designed to optimize some function of the
(traditionally linear) regression model fit; for example, one can seek
to minimize parameter variance or to maximize statistical evidence.
See \cite{AtkiDone1992} for an overview.  \cite{HoiJinLyu2006} provide
an example of optimal design in text analysis.  Our approach in
Section \ref{design} centers on this optimal design literature, and in
particular builds upon {\it D-optimal} designs
\citep[][]{Wald1943,StJoDrap1975} which, if $\bm{X}$ is the sample
covariate matrix, seek to maximize the determinant $|\bm{X}'\bm{X}|$
(thus minimizing the determinant of coefficient covariance for an
ordinary least squares regression onto $\bm{X}$).  When optimal design
is applied to sequential sampling problems, its goals converge with
those of active learning.  The main distinction is that while active
learning is usually focused on adding points one-at-a-time, sequential
optimal design such as in \cite{MullGiov1995} optimizes batch samples.

\subsection{Topic estimation and partial uncertainty quantification}
\label{bayes}

Topic analysis in this article follows the MAP estimation approach of
\citet{Tadd2012b}, yielding jointly optimal $\bs{\Omega}$ and
$\bs{\Theta}$.  Briefly,
parameters are fit to maximize the joint posterior
$L(\bs{\Omega},\bs{\Theta})$ for $\bs{\Omega}$ and $\bs{\Theta}$ {\it
  after} transform into their natural exponential family
parametrization.  This is equivalent to posterior maximization
after adding 1 to each $\alpha$ prior concentration parameter, and is
useful for providing algorithm stability and avoiding boundary
solutions.

For posterior approximation, it is also convenient to work with
document topic weights (i.e., factors) transformed to this natural
exponential family parametrization. That is, $\bs{\Omega}$ is replaced
by $\bs{\Lambda} = \{ \bs{\lambda}_1, \ldots, \bs{\lambda}_n\}$ where for
each document $\omega_k = \exp[{\lambda_{k-1}}]/\sum_{h=0}^{K-1}
\exp[{\lambda_{h}}]$, $k=1\ldots K$, with the fixed element
$\lambda_0=0.$ Given MAP $\bs{\hat \Lambda}$ and $\bs{\hat \Theta}$, a
Laplace approximation \citep[e.g.,][]{TierKada1986} to the posterior
is available as $\mr{p}( \bs{\Lambda},\bs{\Theta} \mid \bm{X}) \approx
\mr{N}\left( \left[\bs{\hat\Lambda},\bs{\hat\Theta}\right], \bm{H}^{-1}
\right),$ where $\bm{H}$ is the log posterior Hessian (i.e., the
posterior information matrix) evaluated at these MAP estimates.  A
further approximation replaces $\bm{H}$ with its block-diagonal,
ignoring off-diagonal elements $\partial^2 L/\partial
\theta_{jk} \partial \lambda_{ih}$.  This allows us to avoid
evaluating and inverting the full matrix $\bm{H}$, a task that is
computationally impractical in large document collections.

The Laplace approximation implies marginal likelihood estimates for a
given $K$, and these are used throughout this article as the basis for
selecting the number of topics.  The approximate posterior also allows
for topic uncertainty quantification through sampling from the {\it
  conditional} posterior for $\bs{\Lambda}$ (or $\bs{\Omega}$) given
$\bs{\Theta}$,
\begin{equation}\label{laplace}
  \mr{p}( \bs{\lambda}_i \mid \bs{\Theta}, \bm{x}_i) \approx
  \mr{N}\left(  \bs{\hat\lambda}_i , \bm{H}_i^{-1} \right),
\end{equation}
where $\bm{H}_i$ has $j^{th}$-row, $k^{th}$-column element $h_{ijk}
= \partial^2L/\partial \lambda_{ij}\partial \lambda_{ik} =
\ds{1}_{[j=k]}\omega_k - \omega_{ij}\omega_{ik}$.  Note that document
factors are independent from each other conditional on $\bs{\Theta}$.
Our approach to posterior approximation is thus to draw
$\bs{\lambda}_{i1} \ldots \bs{\lambda}_{iB}$ from (\ref{laplace}) and
apply the logit transform to obtain $\bs{\omega}_{i1} \ldots
\bs{\omega}_{iB}$ as a sample from $\mr{p}( \bs{\omega}_i \mid
\bs{\Theta}, \bm{x}_i)$.  Although by ignoring uncertainty about
$\bs{\Theta}$ this provides only a partial assessment of variability,
correlation between individual $\bs{\omega}_i$ and $\bs{\Theta}$
decreases with $n$ and the simple normal approximation allows fast
posterior sampling.

 \section*{Acknowledgments}
  Taddy is an Associate Professor of
  Econometrics and Statistics and Neubauer Family Faculty Fellow at
  the University of Chicago Booth School of Business.  The author
  thanks the editor and reviewers, and participants in the
  Department of Energy CoDA workshop for much helpful
  discussion and advice.

%
%

\sgl\small
\bibliographystyle{chicago}
\bibliography{taddy}

\end{document}